\documentclass[aps,prl,tightenlines,twocolumn,floatfix,nofootinbib,showpacs,superscriptaddress]{revtex4}
\usepackage{epsfig}
\usepackage{dcolumn}
\usepackage{graphicx}

\newcommand{\be}{\begin{eqnarray}}
\newcommand{\ee}{\end{eqnarray}}
\newcommand\del{\partial}
\newcommand{\nn}{\nonumber}
\newcommand{\mui}{\mu_{\rm iso}}
\newcommand{\mat}{\left ( \begin{array}{cc}}
\newcommand{\emat}{\end{array} \right )}
\def\bml{\begin{mathletters}}
\def\eml{\end{mathletters}}
\def\ba{\begin{array}}
\def\ea{\end{array}}

\def\nn{\nonumber}
\def\to{\rightarrow}
\def\la{\lambda}
\def\tl{\tilde{\lambda}}
\def\txi{\tilde{\xi}}

\newlength{\bredde}
\def\slash#1{\settowidth{\bredde}{$#1$}\ifmmode\,\raisebox{.15ex}{/}
\hspace*{-\bredde} #1\else$\,\raisebox{.15ex}{/}\hspace*{-\bredde} #1$\fi}
\def\Sl#1{\rlap{\raisebox{.15ex}{$\mskip 4 mu /$}}#1}  

\begin{document}
\title{ A New Method for Determining
$F_\pi$ on the Lattice}
\author{P.H. Damgaard}
\affiliation{The Niels Bohr Institute, Blegdamsvej 17, DK-2100 Copenhagen \O,
Denmark}
\author{Urs M. Heller}
\affiliation{American Physical Society, One Research Road, Box 9000, Ridge,
NY 11961-9000, USA}
\author{K. Splittorff}
\affiliation{NORDITA, Blegdamsvej 17, DK-2100 Copenhagen \O, Denmark}
\author{B. Svetitsky}
\affiliation{School of Physics and Astronomy, Raymond and Beverly Sackler
Faculty of Exact Sciences, Tel Aviv University, 69978 Tel Aviv, Israel
}
\date{\today}
\begin{abstract} 
We derive the two-point spectral correlation function of the Dirac operator 
with a specific external source in the $\epsilon$-regime of QCD.
This correlation function has a unique and strong dependence on $F_\pi$, 
and thus
provides an novel way to extract $F_\pi$ from lattice simulations.
We test the method in a quenched lattice simulation with staggered
fermions.
\end{abstract}
\maketitle


\section{Introduction}
One of the outstanding challenges to lattice gauge theory is the
computation of physical observables that depend strongly on having very 
small quark masses in QCD. 
Here we exploit the fact that the low-lying spectrum of the Dirac operator 
in finite volume is particularly sensitive to the observables of 
spontaneously broken chiral symmetry. 
It is well established~\cite{Jac} that the chiral condensate 
$\Sigma = |\langle\bar{\psi}\psi\rangle|$ can be extracted from 
measurements of the low-energy Dirac spectrum.
We carry this program one step further to determine the next
low-energy constant, the pseudoscalar decay constant $F_{\pi}$, 
with similar high precision.
We show that a certain spectral correlation function of the Dirac operator 
depends on $F_{\pi}$ in a unique and quite spectacular way. 
Based on this dependence we propose and demonstrate a novel method for 
measuring $F_{\pi}$ in lattice gauge theory simulations. The method
is general for systems with spontaneous breaking of symmetries, and
indeed the universal finite-volume scaling formulas have wide application in 
the context of condensed matter physics as well \cite{Efetov}.

Conventionally, measurements of $F_\pi$ on the lattice are carried out in
the so-called $p$-regime using the 2-point function of the
axial current. One then aims at lattices large enough that the Compton 
wavelength of the Goldstone bosons is much smaller than
the lattice size while still performing an extrapolation
to the chiral limit of very light $u$ and $d$ quarks \cite{MILC}.
The method we propose avoids such issues by going to the $\epsilon$-regime. 
 
The low-energy effective Lagrangian of the $\epsilon$-regime in QCD 
\cite{GL,Neuberger} is dominated by the zero-momentum modes of the 
pseudo-Goldstone bosons. In the absence
of external sources, the leading term in the associated $\epsilon$-expansion
is proportional to the quark mass $m$. 
It has been supposed that $F_{\pi}$ needs to be 
computed either from the tiny perturbative correction to the 
leading-order result for $\Sigma$ \cite{D,MT} or from an appropriate 
space-time correlation function \cite{H}. Quenched Monte 
Carlo simulations have demonstrated that such a procedure is feasible, 
but numerically challenging \cite{Berlin}.

Once external currents are included, the leading-order
effective Lagrangian of the $\epsilon$-regime depends
not only on $\Sigma$, but also on $F_{\pi}$.  
We shall introduce an external vector source that 
can be interpreted either as an imaginary chemical potential for
isospin \cite{TV1,CPTmu} or as twisted boundary conditions for
the gauge potentials \cite{MT}. 
The advantage of an {\em imaginary}
isospin chemical potential is twofold.
First, the associated Dirac
operator has a positive definite determinant and
thus becomes amenable to numerical simulations. In addition,  
with an imaginary isospin potential, the Dirac operator is 
anti-Hermitian and thus its
eigenvalue spectrum lies entirely on the imaginary axis. 

For non-zero {\em baryon\/} chemical potential, 
the $F_\pi$-dependent spectral correlation functions in the 
$\epsilon$-regime \cite{O,Kimetal,Kimetal2} can be 
used to glean information about chiral symmetry breaking \cite{OSV}.
However, the spectrum of the associated Dirac operator is complex, 
which makes the numerical determination of $F_\pi$ more demanding.
Quenched results in that direction have recently been presented
\cite{JamesTilo}. 

We consider instead the correlation function 
\be
\rho(\la_1,\tl_2;i\mui) & \equiv & \left\langle \sum_n \delta(\la_1-\la_n)\sum_m
 \delta(\tl_2-\tl_m)\right\rangle \nn \\
-&&\!\!\!\!\!\!\! \left\langle \sum_n \delta(\la_1-\la_n)\!\right\rangle\!\!
         \left\langle\sum_m \delta(\tl_2-\tl_m)\!\right\rangle
\label{corrfdef}
\ee
between the densities of eigenvalues $i\la_n$ of the anti-hermitian 
operator $D_+$, where
\be
D_+\psi_n ~\equiv~ 
[\Sl{D}(A)+i\mui\gamma_0]\psi_n=i\la_n\psi_n, \label{Dplus}
\ee
and the eigenvalues $i\tl_m$ of the likewise anti-hermitian operator
$D_-$, defined by
\be
D_-\tilde{\psi}_n ~\equiv~
[\Sl{D}(A)-i\mui\gamma_0]\tilde{\psi}_n=i\tl_m\tilde{\psi}_n .
\label{Dminus}
\ee
Here $\Sl{D}(A)$ is the Dirac operator associated with the gauge
potential $A_{\mu}$. Considering the operators $D_+$ and $D_-$ as acting on
two separate flavors leads to a theory in which the $u$-quark
has chemical potential $+i\mui$, while the $d$-quark has chemical
potential $-i\mui$. 
Using the effective low energy theory for QCD in the $\epsilon$-regime we
will calculate the correlation function (\ref{corrfdef})
in the microscopic limit where both $m\Sigma V$
and $\mui^2 F_\pi^2V$ are held fixed as the four-volume 
$V$ is taken to infinity. 
We can measure $\Sigma$ very accurately from the distributions 
of the smallest eigenvalues of $\Sl{D}$ \cite{DN}.
Then the spectral correlation function (\ref{corrfdef}) for eigenvalues of
the order $1/(\Sigma V)$ 
provides, as we shall see, a parameter-free determination of 
$F_{\pi}$ in the chiral limit. Computationally, all that is required
is the determination of a modest number (typically, the first 10-20
will suffice) of smallest eigenvalues of the Dirac operators $D_{\pm}$.
As an additional
check, one can verify that the dependence is through the combination
$\mui^2 F_\pi^2V$ only.

In this paper we will provide formulas and
numerical results relevant to quenched QCD. Shortcomings of 
quenching are well known,
and we use this approximation for illustrative purposes only.
The case of dynamical quarks will be presented elsewhere. 
   
\section{The correlation function in the $\epsilon$-regime}

The first step is to derive the quenched susceptibility, defined as
\be
\chi(m_1,m_2;i\mui)\!\equiv\!\lim_{n\to 0} \frac{1}{n^2}\del_{m_1}
\del_{m_2} \log Z_n(m_1,m_2;i\mui)
\label{chi-replica}
\ee
where the limit $n\to0$ indicates use of the replica method \cite{DS}.
In terms of the eigenvalues $\la$ and $\tl$ the susceptibility
is 
\be
\chi(m_1,m_2;i\mui) & = &  \left\langle \sum_n \frac{1}{i\la_n+m_1}\sum_m
 \frac{1}{i\tl_m+m_2}\right\rangle  \\
  &  & \hspace{-4mm}- \left\langle \sum_n \frac{1}{i\la_n+m_1}\right\rangle
         \left\langle\sum_m\frac{1}{i\tl_m+m_2}\right\rangle.\nn
\ee
The spectral correlation function (\ref{corrfdef}) then follows from the 
discontinuity across the imaginary axis of both $m_1$ and~$m_2$,
\be
\rho(\lambda_1,\tl_2;i\mui) \! =\!  \frac1{4\pi^2}
{\rm  Disc}\, \chi(m_1,m_2;i\mui)|_{{m_1=i\la_1}\atop{m_2=i\tl_2}}. 
\label{disc}
\ee
The QCD partition function $Z_n(m_1,m_2;i\mui)$ entering 
Eq.~(\ref{chi-replica})
contains $2n$ fermions, half of which have mass $m_1$ and chemical potential
$i\mui$ while the other half have mass $m_2$ and chemical potential
$-i\mui$. In the $\epsilon$-regime the leading term in 
the partition function is \cite{TV1}
\be\label{zeff}
&&\!\!\!Z_n(m_1,m_2;i\mui) = \\
&&\!\!\!\!\!\!\!\!\!\!
\int\! dU \det(U)^\nu e^{\frac{1}{4}VF_\pi^2\mui^2
{\rm Tr}[U,B][U^\dagger,B] + 
\frac 12 \Sigma V{\rm Tr}(M^\dagger U + MU^\dagger)}\nn
\ee
at fixed gauge field topology $\nu$. The integration is over the Haar
measure of $U(2n)$, and we have defined
\be
\qquad B = \mat \mathbf{1}& 0 \\ 0& -\mathbf{1} \emat \ \ {\rm and} 
\ \ M = \mat m_1\mathbf{1}& 0 \\ 0 & m_2\mathbf{1} \emat.
\ee
For $n=1$ the partition function (\ref{zeff}) reduces to
\be 
&&Z_1(m_1,m_2;i\mui) \\
&&\!\!\!\!=e^{-2VF^2_\pi\mui^2}\!\!\int_0^1\!\! d\lambda\, \lambda\, 
e^{2VF^2_\pi\mui^2\lambda^2} \!
I_\nu(\lambda m_1\Sigma V)I_\nu (\lambda m_2\Sigma V)\nn
\label{Z1mui}
\ee
and from this all partition functions for $n \geq 2$ can be obtained
\cite{Kimetal} via
\be\label{Ztau}
&&\!\!\!\!\!(m_1 m_2)^{n(n-1)}Z_n(m_1,m_2;i\mui)\\
&&= D_n\det \left[(m_1\del_{m_1})^k
  (m_2\del_{m_2})^l Z_1(m_1,m_2;i\mui)\right].\nn 
\ee
Here $D_n$ is a normalization factor and $k,l\!=\!0,1,..,n-1$.  

\begin{figure}[ht]
  \unitlength1.0cm
    \epsfig{file=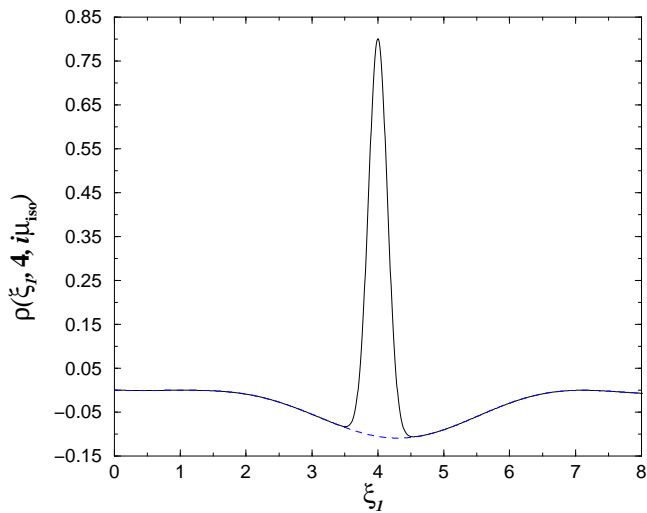,clip=,width=8.5cm}
  \caption{ 
  \label{fig:2pf-1}  
 The correlation function with one eigenvalue fixed at $\xi_2=4$, for 
  $F_\pi\mui\sqrt{V}=0.0717$ (full) and for  $\mu_{\rm iso}=0$
  (dashed). The $\delta$-function peak at $\xi_1=\xi_2$ for  
  $\mui=0$ has not been shown.}
\end{figure}

Recently it has been realized 
\cite{kanzieper,SV} that to obtain the correct replica limit  
$n\to0$ in Eq.~(\ref{chi-replica}) one can
make use of the integrability relations satisfied by the partition functions.
Equation~(\ref{Ztau}) has the structure of a $\tau$-function, implying that 
the $Z_n$ satisfy the Toda lattice
equation,
\be \label{Toda} 
&&\frac{1}{4n^2 V^4\Sigma^4}m_1\del_{m_1} m_2\del_{m_2} 
\log Z_n(m_1,m_2;i\mui) \\
= \!\!\!&&(m_1m_2)^2 \frac
  {Z_{n+1}(m_1,m_2;i\mui)Z_{n-1}(m_1,m_2;i\mui)}
{[Z_n(m_1,m_2;i\mui)]^2} \nn
\ee 
where $D_n$ fixes the coefficient on the lhs \cite{Kimetal}.
Taking the $n\to0$ limit of Eq.~(\ref{Toda}) 
and comparing to Eq.~(\ref{chi-replica}) we find 
\be\label{sucep}
&&\!\!\!\!\!\!\!\!\!\!\!\!\frac{\chi(m_1,m_2;i\mui)}{V^4\Sigma^4}\\
& = & 4m_1m_2 Z_{1}(m_1,m_2;i\mui)Z_{-1}(m_1,m_2;i\mui).\nn
\ee
The $n\to0$ limit in Eq.~(\ref{sucep}) has naturally brought in the
partition function with $n=-1$, i.e.,
one quark of bosonic statistics \cite{SV}.
Like its fermionic analogue, the bosonic partition function is 
determined by the symmetries of the  underlying QCD Lagrangian.
In the bosonic case, moreover, one must take care to ensure  
convergence of the partition function.
As the purely imaginary chemical potential does not affect the 
hermiticity of the Dirac operator this does not lead to additional 
constraints (as opposed to the case of a real chemical potential 
\cite{Kimetal,Kimetal2}).
The result for the bosonic partition function is 
\be
&&Z_{-1}(m_1,m_2;i\mui) =e^{2VF^2_\pi\mui^2} \\
&& \hspace{-4mm} \times \int_1^\infty \!\!d\la \,\la \, 
e^{-2VF^2_\pi\mui^2 \la^2}
K_\nu(\la m_1\Sigma V) K_\nu (\la m_2 \Sigma V).\nn
\label{Z-1}
\ee  
The quenched susceptibility now follows from Eq.~(\ref{sucep}),
\begin{widetext}
\be
\frac{\chi(m_1,m_2;i\mui)}{V^4\Sigma^4} & = & 4m_1 m_2
\int_0^1 d\lambda\, \lambda \,
e^{2VF^2_\pi\mui^2\lambda^2} I_\nu(\lambda m_1 \Sigma V)I_\nu (\lambda
m_2\Sigma V) \nn \\
&&\times   \int_1^\infty d\la\, \la\, 
e^{-2VF^2_\pi\mui^2\la^2} K_\nu(\la m_1 \Sigma V)K_\nu (\la m_2 \Sigma V).
\label{sucep-final}
\ee
Taking the discontinuity as in Eq.~(\ref{disc})
we obtain the desired correlation function, 
\be
\rho(\xi_1,\tilde{\xi}_2;i\mui)
&=& \xi_1\tilde{\xi}_2 \int_0^1 d\la\, \la \,
e^{2VF^2_\pi\mui^2\la^2} J_\nu(\la \xi_1)J_\nu (\la \tilde{\xi}_2) \label{rho2pt}\\[2pt]
&&\times \left[\frac{1}{4 VF^2_\pi\mui^2}
\exp\left({-\frac{\xi_1^2+\tilde{\xi}_2^2}{8 V F^2_\pi \mui^2}}\right)
I_\nu\left(\frac{\xi_1\tilde{\xi}_2}{4 V F^2_\pi\mui^2}\right)\right.\nn 
-\int_0^1 d\la\, \la \,e^{-2VF^2_\pi\mui^2\la^2} 
 J_\nu(\la \xi_1)J_\nu (\la \tilde{\xi}_2)\Bigg] ,
\ee
\end{widetext}
where we have defined the scaling variables 
$\xi_1\equiv\la_1\Sigma V$ and $\tilde{\xi}_2\equiv\tl_2\Sigma V$.
The spectacular change in this correlation 
function when $\mui$ is made non-zero can be seen in Fig.~\ref{fig:2pf-1}.
A fit to Monte Carlo data with $\mui\not=0$ using Eq.~(\ref{rho2pt})
will then readily produce a measurement of $F_\pi$.

The pronounced effect of $\mui\not=0$ in the region where $\la$ is close to
$\tl$ is not difficult to explain. 
For $\mui=0$ both $\la$ and $\tl$ are
eigenvalues of $\Sl{D}$, and 
this leads to a $\delta$-function contribution to the 
correlation function stemming from the non-compact integral 
in the bosonic partition function (\ref{Z-1}), which diverges for $\mui=0$ at
$\la_1=\tl_2$. In this way we recover from Eq.~(\ref{rho2pt})
the known result~\cite{TV},
\be\label{corr-quenched-mu0}
&& \rho(\xi_1,\txi_2;i\mui=0) \\
&&=\delta(\xi_1-\txi_2)\frac{\xi_1}{2}[J_\nu^2(\xi_1)
-J_{\nu+1}(\xi_1)J_{\nu-1}(\xi_1)] \nn\\
&& -\frac{\xi_1\txi_2}{(\xi_1^2-\txi^2)^2}
\left[\xi_1 J_{\nu+1}(\xi_1)J_{\nu}(\txi_2)
    - \txi_2J_{\nu+1}(\txi_2)J_{\nu}(\xi_1)\right]^2.\nn
\ee
The $\delta$-function is not shown in the 
$\mui=0$ curve in Fig.~\ref{fig:2pf-1}. 
When $\mui$ is non-zero, $\la$ is an eigenvalue of $D_+$  
while $\tl$ is an eigenvalue of $D_-$; the effect of $\mui$ is 
therefore to smooth out the $\delta$-function into a pronounced
peak for $\la$ near $\tl$. 


\section{Numerical simulations}

To test the method, we have performed simulations of 
quenched QCD using staggered fermions for
$V=8^4$ ($\mui=0.01$) and $V=12^4$ ($\mui=0.002$). 
We have chosen to work with the 
standard Wilson plaquette action at $\beta=5.7$ and with conventional, 
unimproved Dirac operators. 
In this way we are sure to have no
ambiguities in the identification of the coset space of 
spontaneous chiral symmetry breaking, which here is $U(2n)$. 
The analysis presented above is based on the coset space $SU(2n)$.
We can account for the extra $U(1)$ factor here by setting $\nu=0$ 
in the formulas and comparing to numerical results {\em without} 
fixed topology (see for example the first paper of ref. \cite{D}). 
Simulation at weaker coupling or the use of improved actions and Dirac 
operators will induce a crossover from $\nu=0$ behavior to an explicit 
dependence on topological index \cite{Karietal}, but it is not our
purpose to explore that aspect here.   
 
We include a chemical potential on the lattice in the standard way
\cite{Hasenfratz-Karsch}. For an
imaginary chemical potential this amounts to including a constant
abelian gauge field with non-vanishing timelike component $\mu$
only. It leaves the hermiticity properties of the Dirac operator
unchanged, allowing the computation of the lowest-lying eigenvalues
with the Ritz variational algorithm \cite{Kalkreuter:1995mm}.

When extracting physical observables we must keep in
mind that the continuum theory describes 4 tastes of
quark. In our simulation the staggered Dirac matrix is 4 times 
larger than that of a single continuum quark. The statistical properties of
the eigenvalue spectrum of that matrix behave as in a theory of one
species in a volume four times as big. Therefore, to determine the 
values of $F_\pi$ and $\Sigma$ from the staggered eigenvalue spectrum 
we replace $V$ in the analytical predictions by 4$V$.

\begin{figure}[ht]
  \unitlength1.0cm
    \epsfig{file=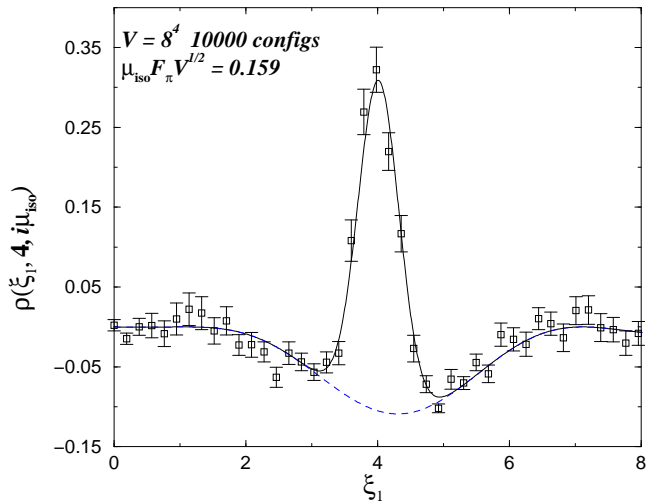,clip=,width=8.5cm}
  \caption{ 
  \label{fig:corr-noint}  
  The correlation function with fixed $\tilde{\xi}_2=4$ measured on our $8^4$
  lattice. The curves correspond to (\ref{rho2pt}) with $F_\pi\mui
  \sqrt{V}=0.159$ (full) and $\mui=0$ (dashed).}
\end{figure}

First we measure $\Sigma$. We can do that by fitting individual eigenvalue
distributions to the analytical expressions \cite{DN}. It is a non-trivial
prediction that $k$-point correlation functions of $D_+$ and $D_-$
separately are independent of $\mui$ in the microscopic limit. This
follows from Eq.~(\ref{zeff}) by taking $B$ proportional to the unit matrix,
and it ensures in particular that individual eigenvalue distributions are
$\mui$-independent in this limit \cite{AD}. 
Alternatively, one can use the 2-point correlation functions of either $D_+$
or $D_-$. As follows from the argument above, these 2-point functions
are $\mui$-independent, and they can provide independent
determinations of $\Sigma$ \cite{Tilo}. Using the latter approach for
our $12^4$ data, a best fit gives in lattice units 
the bare value $\Sigma = 0.0634(3)$ with $\chi^2/dof = 0.55$. This
is consistent with what we find by fitting either individual
eigenvalue distributions or the overall flat eigenvalue plateau. 
We can then determine $F_\pi$ by a fit of the measured correlation 
function (\ref{corrfdef}) to the analytical result (\ref{rho2pt}).
In Fig. \ref{fig:corr-noint} we show the data for $V=8^4$ and compare
them to the analytical curves using the best fit to $F_{\pi}$ from the $12^4$
data, as explained below. 
On $V=8^4$ with $\beta=5.7$ only the first few eigenvalues have
distributions in agreement with the predictions of the
$\epsilon$-regime. For this reason we base our measured values of
$\Sigma$ and $F_\pi$ on our $12^4$ lattice simulations.

\begin{figure}[ht]
  \unitlength1.0cm
    \epsfig{file=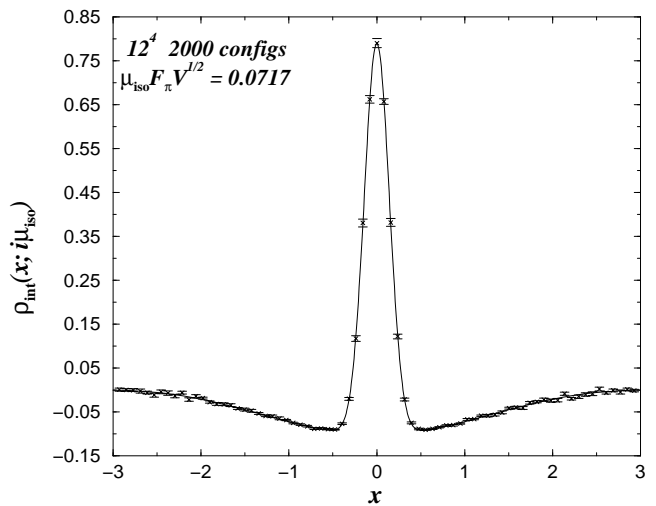,clip=,width=8.5cm}
  \caption{ 
  \label{fig:2pf-2}  
  The integrated correlation function (\ref{corr-int}) 
  with $\txi_{\rm  min}=4$ and $\txi_{\rm max}=43$ . 
  The value $F_\pi\mui \sqrt{V}=0.0717$ is the best fit.}
\end{figure}

Requiring that the eigenvalue $\tilde{\xi}_2$ fall within one bin
 around a fixed value (here $\tilde{\xi}_2=4$) means that a large 
fraction of the lattice configurations is obviously not used. 
In order to improve the statistics we consider the 
integrated correlation function
\be\label{corr-int}
\rho_{\rm int}(x;i\mui)
\equiv\int_{\txi_{\rm min}}^{\txi_{\rm max}}{\rm d}\txi\, 
\rho(x+\txi,\txi;i\mui).
\ee
In Fig.~\ref{fig:2pf-2} we show this integrated correlation function as
measured on our $12^4$ ensemble.

The best fit gives us $F_\pi =0.1245(18)$ in bare lattice units with a 
$\chi^2/dof =0.33$. This value is consistent with the result $F_\pi=0.118(7)$ 
of ref.~\cite{Gupta}, which uses the same action and gauge coupling 
$\beta=5.7$ (but of course a different method for extracting
$F_\pi$). 

It is well known that the unimproved staggered action leads to serious
scaling violations. To quote a result in physical units would, at this
value of $\beta$, require the use of improvement methods to reduce
lattice artifacts. This is beyond the scope of this paper.


\section{Summary}

We have proposed and tested a new
method to measure $F_{\pi}$ in lattice gauge theory 
simulations, using the pronounced $F_{\pi}$-dependence
of a specific 2-point correlation function of the Dirac
operator in the $\epsilon$-regime. 

We have performed quenched lattice simulations that demonstrate 
excellent agreement with our analytical predictions. Our study
illustrates the ease with which this method can be implemented, 
and the high precision that can be achieved.

\vspace{2mm}
{\em Acknowledgement:}~
PHD, UMH and KS would like to thank the Kavli Institute of Theoretical
Physics for its hospitality and NSF grant PHY99-07949 for partial
support.
The work of BS was supported in part by the Israel Science Foundation 
under grant no.~222/02-1.
The bulk of our numerical simulations were carried out on an 
SGI Origin~2800 computer operated by the 
High Performance Computing Unit of the Israel Inter-University Computation 
Center. Our computer code is based on the public lattice gauge theory
code of the MILC Collaboration \cite{MILC1}.


\begin{thebibliography}{0}

\bibitem{Jac}E.~V.~Shuryak and J.~J.~M.~Verbaarschot,
  Nucl.\ Phys.\ A {\bf 560}, 306 (1993)
  [hep-th/9212088];
M.~E.~Berbenni-Bitsch et al.,
  Nucl.\ Phys.\ Proc.\ Suppl.\  {\bf 63}, 820 (1998)
  [hep-lat/9709102];
P.~H.~Damgaard, U.~M.~Heller and A.~Krasnitz,
  Phys.\ Lett.\ B {\bf 445}, 366 (1999)
  [hep-lat/9810060].

\bibitem{Efetov} See $e.g.$ K.~B.~Efetov, {\em Supersymmetry in Disorder
and Chaos}, Cambridge Univ. Press (1997). 

\bibitem{MILC} For a recent study see
C.~Aubin {\it et al.},
  Phys.\ Rev.\ D {\bf 70}, 114501 (2004)
  [hep-lat/0407028].


\bibitem{GL}J.~Gasser and H.~Leutwyler,
  Phys.\ Lett.\ B {\bf 184}, 83 (1987);
  Phys.\ Lett.\ B {\bf 188}, 477 (1987);
H.~Leutwyler and A.~Smilga,
  Phys.\ Rev.\ D {\bf 46}, 5607 (1992).


\bibitem{Neuberger}
H.~Neuberger,
  Phys.\ Rev.\ Lett.\  {\bf 60}, 889 (1988);
  Nucl.\ Phys.\ B {\bf 300}, 180 (1988).


\bibitem{D}
P.~H.~Damgaard,
Nucl.\ Phys.\ B {\bf 608}, 162 (2001)
[hep-lat/0105010];
P.~H.~Damgaard, M.~C.~Diamantini, P.~Hernandez and K.~Jansen,
  Nucl.\ Phys.\ B {\bf 629}, 445 (2002)
  [hep-lat/0112016].

\bibitem{MT}
T.~Mehen and B.~C.~Tiburzi,
hep-lat/0505014.

\bibitem{H}
F.~C.~Hansen,
  Nucl.\ Phys.\ B {\bf 345}, 685 (1990);
P.~H.~Damgaard et al.,
  Nucl.\ Phys.\ B {\bf 656}, 226 (2003)
  [hep-lat/0211020].

\bibitem{Berlin}
W.~Bietenholz et al.,
  JHEP {\bf 0402}, 023 (2004)
  [hep-lat/0311012];
L.~Giusti et al.
  JHEP {\bf 0404}, 013 (2004)
  [hep-lat/0402002];
H.~Fukaya, S.~Hashimoto and K.~Ogawa,
  hep-lat/0504018.

\bibitem{CPTmu}
  J.~B.~Kogut, M.~A.~Stephanov and D.~Toublan,
  Phys.\ Lett.\ B {\bf 464}, 183 (1999)
  [hep-ph/9906346];
  D.~T.~Son and M.~A.~Stephanov,
  Phys.\ Rev.\ Lett.\  {\bf 86}, 592 (2001)
  [hep-ph/0005225];
  K.~Splittorff, D.~Toublan and J.~J.~M.~Verbaarschot,
  Nucl.\ Phys.\ B {\bf 620}, 290 (2002)
  [hep-ph/0108040];
  Nucl.\ Phys.\ B {\bf 639}, 524 (2002)
  [hep-ph/0204076].


\bibitem{TV1}
D.~Toublan and J.~J.~M.~Verbaarschot,
  Int.\ J.\ Mod.\ Phys.\ B {\bf 15}, 1404 (2001)
  [hep-th/0001110].
\bibitem{O}  J. C. Osborn, Phys. Rev. Lett. {\bf 93}, 222001 (2004).

\bibitem{Kimetal}
K.~Splittorff and J.~J.~M.~Verbaarschot,
  Nucl.\ Phys.\ B {\bf 683}, 467 (2004)
  [hep-th/0310271].
  
\bibitem{Kimetal2}  
G.~Akemann, J.~C.~Osborn, K.~Splittorff and J.~J.~M.~Verbaarschot,
  Nucl.\ Phys.\ B {\bf 712}, 287 (2005)
  [hep-th/0411030].

\bibitem{OSV}
  J.~C.~Osborn, K.~Splittorff and J.~J.~M.~Verbaarschot,
  Phys.\ Rev.\ Lett.\  {\bf 94} (2005) 202001
  [hep-th/0501210].

\bibitem{JamesTilo} J. Osborn and T. Wettig, talk at XQCD in Swansea (2005).

\bibitem{DN}
P.~H.~Damgaard and S.~M.~Nishigaki,
  Phys.\ Rev.\ D {\bf 63}, 045012 (2001)
  [hep-th/0006111].


\bibitem{DS}
See for example: P.~H.~Damgaard and K.~Splittorff,
  Phys.\ Rev.\ D {\bf 62}, 054509 (2000)
  [hep-lat/0003017].



\bibitem{kanzieper} E.~Kanzieper,
  Phys.\ Rev.\ Lett.\  {\bf 89}, 250201 (2002)
  [cond-mat/0207745].

\bibitem{SV}
K.~Splittorff and J.~J.~M.~Verbaarschot,
  Phys.\ Rev.\ Lett.\  {\bf 90}, 041601 (2003)
  [cond-mat/0209594];
  Nucl.\ Phys.\ B {\bf 695} (2004) 84
  [hep-th/0402177].


\bibitem{AD}
G.~Akemann and P.~H.~Damgaard,
  Phys.\ Lett.\ B {\bf 583}, 199 (2004)
  [hep-th/0311171].

\bibitem{TV}
A.~V.~Andreev, B.~D.~Simons and N.~Taniguchi,
Nucl.\ Phys.\ B {\bf 432} 487 (1994);
D.~Toublan and J.~J.~M.~Verbaarschot,
  Nucl.\ Phys.\ B {\bf 603}, 343 (2001)
  [hep-th/0012144].

\bibitem{Hasenfratz-Karsch}
  P.~Hasenfratz and F.~Karsch,
  Phys.\ Lett.\ B {\bf 125}, 308 (1983).

\bibitem{Kalkreuter:1995mm}
  T.~Kalkreuter and H.~Simma,
  Comput.\ Phys.\ Commun.\  {\bf 93}, 33 (1996)
  [hep-lat/9507023].


\bibitem{Karietal}
P.~H.~Damgaard, U.~M.~Heller, R.~Niclasen and K.~Rummukainen,
  Phys.\ Rev.\ D {\bf 61}, 014501 (2000)
  [hep-lat/9907019];
E.~Follana, A.~Hart and C.~T.~H.~Davies,
  Phys.\ Rev.\ Lett.\  {\bf 93}, 241601 (2004)
  [hep-lat/0406010];
S.~Durr, C.~Hoelbling and U.~Wenger,
  Phys.\ Rev.\ D {\bf 70}, 094502 (2004)
  [hep-lat/0406027];
K.~Y.~Wong and R.~M.~Woloshyn,
  Phys.\ Rev.\ D {\bf 71}, 094508 (2005)
  [hep-lat/0412001];
E.~Follana, A.~Hart, C.~T.~H.~Davies and Q.~Mason,
  hep-lat/0507011.



\bibitem{Tilo}
The
2-point function for $\mui=0$
was first measured by 
M.~E.~Berbenni-Bitsch et al.,
  Phys.\ Rev.\ Lett.\  {\bf 80}, 1146 (1998)
  [hep-lat/9704018].

\bibitem{Gupta}
  R.~Gupta, G.~Guralnik, G.~W.~Kilcup and S.~R.~Sharpe,
  Phys.\ Rev.\ D {\bf 43}, 2003 (1991).

\bibitem{MILC1}
        Available from http://www.physics.utah.edu/$^\sim$detar/milc/



\end{thebibliography}
\end{document}